\documentstyle[sprocl]{article}

\def\D0bar{\overline D{}^0}
\def\barD{\overline D{}^0}

\def\DG{\Delta \Gamma_D}

\def\be{\begin{equation}}
\def\ee{\end{equation}}
\def\bea{\begin{eqnarray}}
\def\eea{\end{eqnarray}}

\begin{document}

\title{WIDTH DIFFERENCE IN THE $D^0-\D0bar$ SYSTEM\footnote{To be 
published in the proceedings of 4th Workshop on Continuous 
Advances in QCD, Minneapolis, Minnesota, 12-14 May 2000. }}

\author{ALEXEY A. PETROV}
\address{~}
\address{Department of Physics and Astronomy, 
	 The Johns Hopkins University,\\
         3400 North Charles Street, Baltimore, Maryland 21218 USA}
\address{and}
\address{Laboratory of Nuclear Studies, Cornell University,\\
         Ithaca, New York 14853 USA\footnote{after September 1st 2000}\\
  	E-mail: petrov@mail.lns.cornell.edu}


\maketitle\abstracts{
The motivation most often cited in searches
for $D^0 - {\bar D}^0$ mixing lies with the possibility of
observing a signal from new physics which dominates that
from the Standard Model.
We discuss recent theoretical and experimental results in 
$D^0 - \barD$ mixing, including new experimental measurements 
from CLEO and FOCUS collaborations and their interpretations.}
\section{Introduction}

Neutral meson-antimeson mixing provides important
information about electroweak symmetry breaking and quark
dynamics. In that respect, the $D^0-\D0bar$ system is unique 
as it is the only system that is sensitive to the dynamics 
of the bottom-type quarks. The
$D^0-\D0bar$ mixing proceeds extremely slowly, which in the Standard
Model (SM)~\cite{Ge92,Oh93,ap,Da85,dght} is usually attributed
to the absence of superheavy quarks destroying GIM cancelations.
This feature makes it sensitive both to physics beyond the 
Standard Model and to long-distance QCD effects.

The low energy effect of new physics particles can be naturally 
written in terms of a series of local operators of increasing 
dimension generating $\Delta C = 2$ transitions. These operators, 
along with the Standard Model contributions, generate the mass
and width splittings for the eigenstates of $D^0-\D0bar$ mixing
matrix defined as
\begin{equation} \label{definition1}
| D_{^1_2} \rangle = 
p | D^0 \rangle \pm q | \bar D^0 \rangle \ \ ,
\end{equation}
with complex parameters $p$ and $q$ determined from
the phenomenological (CPT-invariant) $D^0-\D0bar$ mass 
matrix~\cite{Donoghue:1992dd}.
It is convenient to normalize the mass and width differences  
to define two dimensionless variables $x$ and $y$ 
\begin{eqnarray} \label{definition}
x \equiv \frac{m_2-m_1}{\Gamma}, ~~
y \equiv \frac{\Gamma_2 - \Gamma_1}{2 \Gamma}.
\end{eqnarray}
where $m_i(\Gamma_i)$ is a mass (width) of the 
corresponding state, $D_{^1_2}$. Clearly,
$y$ is built from the decays of $D$ into the physical 
states, and so it should be dominated by the SM contributions.
If CP-violation is neglected, then
$p=q$ and $| D_{^1_2} \rangle$ become eigenstates of $CP$.
To set up a relevant formalism, let us recall that in
perturbation theory, the $ij^{\rm th}$ element of the
$D^0 - \D0bar$ mass matrix can be represented as
\begin{equation}
\left [ M - i \frac{\Gamma}{2} \right ]_{ij} =
\frac{1}{2 m_D}
\langle D^0_i | {\cal H}^{\Delta C=2}_W | D^0_j \rangle
+ \frac{1}{2 m_D} \sum_{I}
\frac{\langle D^0_i | {\cal H}^{\Delta C=1}_W | I \rangle
\langle I | {\cal H}_W^{\Delta C=1 \dagger} | D^0_j \rangle}
{m_D^2 - E_I^2 + i \epsilon} \ \ .
\label{mixmatr}
\end{equation}
Here the first term of Eq.~(\ref{mixmatr}) comes from the local
$\Delta C = 2$ (box and dipenguin) operators.  These contributions
affect $\Delta M$ only and expected to be small in the Standard
Model~\cite{ap,Da85,dght}. It is therefore natural to expect that the
$\Delta C = 2$ part of Eq.~(\ref{mixmatr}) might receive contributions
from the effective operators generated by the new physics
interactions. Next come the bilocal contributions
which are induced by the insertion of two Hamiltonians
changing the charm quantum number by one unit, i.e. built out of 
$\Delta C=1$ operators.
This class of terms contributes to both $x$ and
$y$ and is believed to give the dominant SM contribution
to the mixing due to various nonperturbative effects. 
Some enhancement due to the $\Delta C =1$
operators induced by new physics is also possible, but unlikely
given the strong experimental constraints provided by the
data on $D$ meson decays. Yet,  
the motivation most often cited in searches
for $D^0 - {\bar D}^0$ mixing lies with the possibility of
observing a signal from new physics which dominates that
from the Standard Model. It is therefore extremely important 
to estimate the Standard Model contribution to $x$ and $y$. 

The mass and width differences $x$ and $y$ can be measured in a
variety of ways, for instance in semileptonic $D \to K l \nu$
or nonleptonic $D \to KK$ or $D \to K\pi$ decays. Let us define
the $D$ meson decay amplitudes into a final state $f$ as
\begin{equation}
A_f \equiv \langle f|{\cal H}_W^{\Delta C=1}|D^0 \rangle,~~ 
\bar A_f \equiv \langle f|{\cal H}_W^{\Delta C=1}|\barD \rangle.
\end{equation}
It is also useful to define the complex parameter $\lambda_f$:
\be
\lambda_f \equiv {q\over p}\ {\bar A_f\over A_f}.
\ee
Let us first consider the processes that are relevant to the 
FOCUS~\cite{Link:2000cu} and CLEO~\cite{Godang:1999yd} experiments.
Those are the doubly-Cabibbo-suppressed $D^0\rightarrow K^+\pi^-$ decay,
the singly-Cabibbo-suppressed $D^0\rightarrow K^+K^-$ decay,
the Cabibbo-favored $D^0\rightarrow K^-\pi^+$ decay,
and the three CP-conjugate decay processes. Let us write down approximate 
expressions for the time-dependent decay rates that are valid for times 
$t < 1/\Gamma$. We take into account the experimental information that $x$, 
$y$ and $\tan\theta_c$ are small, and expand each of the rates only to the
order that is relevant to the CLEO and FOCUS measurements:
\begin{eqnarray} \label{cleofocus}
&& \Gamma[D^0(t)\rightarrow \ K^+\pi^-]\ = \ e^{-\Gamma t}|
\bar A_{K^+\pi^-}|^2 |q/p|^2 
\nonumber \\ 
&&~~~~~~\times \left\{|\lambda_{K^+\pi^-}^{-1}|^2+
[\Re(\lambda^{-1}_{K^+\pi^-})y+\Im(\lambda^{-1}_{K^+\pi^-})x]\Gamma t
+{1\over4}(y^2+x^2)(\Gamma t)^2\right\}, 
\nonumber \\
&& \Gamma[\barD(t)\rightarrow\ K^-\pi^+]\ =\ 
e^{-\Gamma t}|A_{K^-\pi^+}|^2 |p/q|^2 
\nonumber \\
&&~~~~~~\times \left\{|\lambda_{K^-\pi^+}|^2+[\Re(\lambda_{K^-\pi^+})y
+\Im(\lambda_{K^-\pi^+})x]\Gamma t+{1\over4}(y^2+x^2)(\Gamma t)^2\right\}, 
\\
&& \Gamma[D^0(t)\rightarrow K^+K^-]\ =\ e^{-\Gamma t}|A_{K^+K^-}|^2
\left\{1+[\Re(\lambda_{K^+K^-})y-\Im(\lambda_{K^+K^-})x]\Gamma t\right\},
\nonumber \\
&& \Gamma[\barD(t)\rightarrow K^+K^-]\ =\ e^{-\Gamma t}|\bar A_{K^+K^-}|^2
\left\{1+[\Re(\lambda^{-1}_{K^+K^-})y-\Im(\lambda^{-1}_{K^+K^-})x]
\Gamma t\right\},
\nonumber \\
&& \Gamma[D^0(t)\rightarrow \ K^-\pi^+]\ =\ e^{-\Gamma t}|A_{K^-\pi^+}|^2,
~\Gamma[\barD(t)\rightarrow \ K^+\pi^-]\ =\ 
e^{-\Gamma t}|\bar A_{K^+\pi^-}|^2.
\nonumber
\end{eqnarray}
Within the Standard Model, the physics of $D^0 - \D0bar$ mixing and of the
tree level decays is dominated by the first two generations and,
consequently, CP violation can be safely neglected. In all `reasonable'
extensions of the Standard Model, the six decay modes of 
Eq.~(\ref{cleofocus}),
are still dominated by the Standard Model CP conserving 
contributions. On the other hand, there could be new short distance, 
possibly CP violating contributions to the mixing amplitude $M_{12}$.
Allowing for only such effects of new physics, the picture of CP violation
is simplified since there is no direct CP violation. The effects of indirect
CP violation can be parameterized in the following way
\begin{eqnarray} \label{defphases}
|q/p|\ &=&\ R_m,\nonumber \\
\lambda^{-1}_{K^+\pi^-}\ &=&\ \sqrt{R}\ R_m^{-1}\ e^{-i(\delta+\phi)},
\nonumber \\
\lambda_{K^-\pi^+}\ &=&\ \sqrt{R}\ R_m\ e^{-i(\delta-\phi)},
\\
\lambda_{K^+K^-}\ &=&\ -R_m\ e^{i\phi}.
\nonumber
\end{eqnarray}
Here $R$ and $R_m$ are real and positive dimensionless numbers.
CP violation in mixing is related to $R_m\neq1$ while CP violation in the 
interference of decays with and without mixing is related to $\sin\phi\neq0$. 
The choice of phases and signs in Eq.~(\ref{defphases}) is consistent 
with having the weak phase difference $\phi=0$ in the Standard Model 
and the strong phase difference $\delta=0$ in the $SU(3)$ limit. 
The weak phase $\phi$ is universal for $K\pi$ and $KK$ final states
under our assumption of negligible direct CP violation.
We further define
\begin{eqnarray} \label{defphases1}
x^\prime\ &\equiv&\ x\cos\delta+y\sin\delta,
\nonumber \\
y^\prime\ &\equiv&\ y\cos\delta-x\sin\delta.
\end{eqnarray}
With the assumption that there is no direct CP violation in the processes
that we study, and using the parameterizations (\ref{defphases})
and (\ref{defphases1}), we can rewrite Eqs.~(\ref{cleofocus}) as 
follows:
\begin{eqnarray}
&& \Gamma[D^0(t)\rightarrow\ K^+\pi^-]\ =\ e^{-\Gamma t}|A_{K^-\pi^+}|^2
\nonumber \\
&&~~~~~~\times~ \left[R+\sqrt{R}R_m(y^\prime\cos\phi-
x^\prime\sin\phi)\Gamma t +{R_m^2\over4}(y^2+x^2)(\Gamma t)^2\right],
\nonumber \\
&&\Gamma[\barD(t)\rightarrow\ K^-\pi^+]\ =\ e^{-\Gamma t}|A_{K^-\pi^+}|^2
\nonumber \\
&&~~~~~~\times~ \left[R+\sqrt{R}R_m^{-1}(y^\prime\cos\phi+x^\prime\sin\phi)
 \Gamma t+ {R_m^{-2}\over4}(y^2+x^2)(\Gamma t)^2\right] 
\nonumber \\
&& \Gamma[D^0(t)\rightarrow\ K^+K^-]\ = \ e^{-\Gamma t}|A_{K^+K^-}|^2
\left[1-R_m(y\cos\phi-x\sin\phi)\Gamma t\right],
\\
&& \Gamma[\barD(t)\rightarrow \ K^+K^-]\ = \ e^{-\Gamma t}|A_{K^+K^-}|^2
\left[1-R_m^{-1}(y\cos\phi+x\sin\phi)\Gamma t\right],
\nonumber \\
&& \Gamma[D^0(t)\rightarrow K^-\pi^+]\ =\ 
\Gamma[\barD(t)\rightarrow K^+\pi^-]\ =\ e^{-\Gamma t}|A_{K^-\pi^+}|^2.
\nonumber
\end{eqnarray}
By studying various combinations of these modes we can
pin down the values of $x$ and $y$ in $D^0-\barD$ system.

\section{Theoretical expectations}

The leading piece of the short-distance part of the mixing 
amplitude is known to be small~\cite{Ge92,Oh93,ap,Da85,dght}, but
it is instructive to see why it is so. We will also complement the
discussion by including leading $1/m_c$ corrections.

As discussed above, the lifetime difference is associated with 
the long-distance contribution to Eq.~(\ref{mixmatr}), i.e. the
double insertion of $\Delta C =1$ effective Hamiltonian
\begin{eqnarray} \label{hamiltonian}
{\cal H}^{\Delta C=1}_W =
- \frac{G_F}{\sqrt{2}} \sum_q \xi_q
\left \{ C_1 (\mu) \bar u_\alpha \Gamma_\mu q_\beta
\bar q_\beta \Gamma^\mu c_\alpha +
C_2 (\mu) \bar u_\alpha \Gamma_\mu q_\alpha
\bar q_\beta \Gamma^\mu c_\beta \right \} 
\end{eqnarray}
where $ \Gamma_\mu = \gamma_\mu (1+\gamma_5)$ and
$\xi_q=V_{cq}^* V_{uq}$ represents the appropriate
CKM factor for $\psi = d,s$. $C_1(m_c) \simeq -0.514$ and
$C_2(m_c) \simeq 1.270$, as found in a NLO QCD calculation
with `scheme-independent' prescription. Hereafter we shall 
not write the scale dependence of Wilson coefficients explicitly.
The width difference $y$ can be written as an imaginary part
of the matrix element of the time-ordered product of 
two $\Delta C  = 1$ Hamiltonians of Eq.~(\ref{hamiltonian}).
Physically, it is generated by a set of on-shell 
intermediate states, and therefore, constitutes an intrinsically
non-local quantity. However, in the limit 
$m_c/\Lambda_{QCD} \to \infty$ the momentum 
flowing through the light ($s$ and $d$ quark) degrees of freedom 
is large and an Operator Product Expansion (OPE) can be performed. 
As a result, both $x$ and $y$ can be represented by a series of 
matrix elements of {\it local} operators of increasing dimension.
In other words, if a typical hadronic distance $z \gg 1/m_c$,
then the decay is a local process.
Of course, significant corrections to the leading term of 
this series are expected, as the expansion parameter 
$\Lambda/m_c$ ($\Lambda \sim \Lambda_{QCD}$ is some hadronic 
parameter) is not small.

It is well known that $y$ should vanish in the limit of equal 
quark masses by the virtue of GIM cancelation mechanism. For the 
$D\bar D$ system it is equivalent to the requirement of flavor 
$SU(3)$ symmetry. The question here is by how much $SU(3)$ is broken.
The (parametrically) leading contribution to $x$ and $y$
comes from the matrix elements of operators of dimension six
\begin{eqnarray}
O_1 &=& \bar u \gamma_\mu (1 + \gamma_5) c
 \bar u \gamma_\mu (1 + \gamma_5) c, ~~~~~~~~
O_1'=\bar u (1 - \gamma_5) c \bar
u  (1 - \gamma_5) c \nonumber \\
O_2 &=& \bar u_i \gamma_\mu (1 + \gamma_5) c_k
 \bar u_k \gamma_\mu (1 + \gamma_5) c_i, ~~~~
O_2'=\bar u_i (1 - \gamma_5) c_k \bar
u_k  (1 - \gamma_5) c_i ~~~
\end{eqnarray}
Using Fierz identities and performing necessary integrations
we obtain
\begin{eqnarray} \label{gsd}
&& ~\Delta \Gamma_D^{(6)} = \frac{N_c+1}{\pi N_c} X_D
{(m_s^2 - m_d^2)^2 \over m_c^2} {m_s^2 + m_d^2 \over m_c^2}
\Biggl[ C_2^2 + 2C_1 C_2 + C_1^2 N_C
\nonumber \\
&&~~-~ 
\frac{2(2N_c-1)}{1+N_c} {B_D' \over B_D} {M_D^2\over (m_c + m_u)^2}
\left(C_2^2 +
\frac{2-N_c}{2 N_c-1}\left( C_1^2 N_c + 2 C_1 C_2 \right) \right)
\Biggr],
\end{eqnarray}
with $N_c=3$ being the number of colors.
This result was reported in~\cite{GPunp}. Numerically, the effect of
including QCD evolution amounts to the enhancement of the
box diagram estimate by approximately a factor of two.
As one can easily see, a standard box diagram contribution is
recovered in the limit $C_1 \to 0, ~C_2 \to 1$ where the QCD 
evolution is turned off
\begin{eqnarray} \label{msd}
\Delta m_D^{\rm box} &=& {2 \over 3\pi^2} X_D
{(m_s^2 - m_d^2)^2 \over m_c^2} \left[
1 - {5\over 4} {B_D' \over B_D} {M_D^2\over (m_c + m_u)^2} \right],
\nonumber \\
\Delta \Gamma_D^{\rm box} &=& {4 \over 3\pi} X_D
{(m_s^2 - m_d^2)^2 \over m_c^2} {m_s^2 + m_d^2 \over m_c^2}
\left[ 1 - {5\over 2} {B_D' \over B_D} {M_D^2\over (m_c + m_u)^2}
\right] \ ,
\end{eqnarray}
with $X_D$ is given by $X_D \equiv \xi_s \xi_d B_D G_F^2 M_D F_D^2$.
Also, the B-parameters $B_D=B_D'=1$ in the usual vacuum saturation 
approximation to
\begin{eqnarray}
\langle D^0 | O_1 | \bar D^0 \rangle &=&
\left(1 + \frac{1}{N_c} \right) \frac{4 F_D^2 m_D^2}{2 m_D} B_D,
\nonumber \\
\langle D^0 | O_1' | \bar D^0 \rangle &=&
-\left(1 - \frac{1}{2 N_c} \right) \frac{4 m_D^2}{(m_c+m_u)^2}
\frac{F_D^2 m_D^2}{2 m_D} B_D', \\
\langle D^0 | O_2 | \bar D^0 \rangle &=&
\left(1 + \frac{1}{N_c} \right) \frac{4 F_D^2 m_D^2}{2 m_D} B_D,
\nonumber \\
\langle D^0 | O_2' | \bar D^0 \rangle &=&
-\left(\frac{1}{N_c} - \frac{1}{2} \right) \frac{4 m_D^2}{(m_c+m_u)^2}
\frac{F_D^2 m_D^2}{2 m_D} B_D',
\nonumber
\end{eqnarray}
where $2 m_D$ in the denominator comes from the normalization
of meson states and $F_D$ is a $D$-meson decay constant.
It is clear from Eq.~(\ref{gsd}) that the smallness of the
leading order result comes from the factor of
$(m_s^2-m_d^2)^2/m_c^2$ which represents the GIM cancelation
among the intermediate $s$ and $d$ quark states and from the
factor $(m_s^2+m_d^2)/m_c^2$ which represents the
helicity suppression of the intermediate state quarks.
At the end, $y \ll x \ll 0.1 \%$.

Of course, one should be concerned with the size of
(parametrically suppressed) corrections to Eq.~(\ref{gsd}). 
This is especially important for the calculation of 
$y$ because of the $SU(3)$ and helicity suppression
of the parametrically leading term. For example,
perturbative QCD corrections, while
suppressed by $\alpha_s(m_s)$, include the gluon
emission diagrams, which do not exhibit helicity 
supression factors of $m_s^2$.

In addition, both $SU(3)$ and helicity suppression factors
$(m_s^2-m_d^2)^2 (m_s^2+m_d^2)$ can be lifted at higher orders 
in $\Lambda/m_c$, which calls for a certain reorganization 
of the operator expansion. In spite of being parametrically 
suppressed, those ``corrections'' are in fact numerically 
larger then the leading order term. It was 
realized~\cite{Ge92,Bigi:2000wn} 
that the higher order contributions from the operators of
dimension nine and twelve that represent interactions 
with the background quark condensates do exactly that.

Taking into account new operator structures generated by the 
renormalization group running of the effective Hamiltonian 
from $M_W$ down to $m_c$, the contribution of dimension nine 
operators reads
\begin{eqnarray}
&& \Delta M_D^{(9)} = 4 \xi_s^2 G_F^2 \frac{m_s^2-m_d^2}{N_c m_c^3} 
v^\alpha \Biggl \{ 
(N_c C_1^2 + 2 C_1 C_2 + C_2^2) 
\nonumber \\
&\times& \biggl[
\langle D^0 | (\bar u \Gamma_\alpha c)(\bar u \Gamma_\mu c)
(\bar \psi \Gamma^\mu \psi) | \D0bar \rangle + \mbox{others} \biggr]
\nonumber \\
&+& 2 C_2^2  \biggl[
\langle D^0 | (\bar u \Gamma_\alpha T^a c)(\bar u \Gamma_\mu T^a c)
(\bar \psi \Gamma^\mu \psi) | \D0bar \rangle +
\langle D^0 | (\bar u \Gamma_\alpha c)(\bar u \Gamma_\mu T^a c)
(\bar \psi \Gamma^\mu T^a \psi) | \D0bar \rangle 
\nonumber \\
&+&~~~ \langle D^0 | (\bar u \Gamma_\alpha T^a c)(\bar u \Gamma_\mu c)
(\bar \psi \Gamma^\mu T^a \psi) | \D0bar \rangle + \mbox{others} \biggr]
\\
&+& 2 N_c C_2^2 (d^{abc} + i f^{abc}) \biggl[
\langle D^0 | (\bar u \Gamma_\alpha T^a c)(\bar u \Gamma_\mu T^b c)
(\bar \psi \Gamma^\mu T^c \psi) | \D0bar \rangle + \mbox{others} \biggr]
\nonumber \\
&+& 4 N_c C_1 C_2 \biggl[
\langle D^0 | (\bar u \Gamma_\alpha c)(\bar u \Gamma_\mu T^a c)
(\bar \psi \Gamma^\mu T^a \psi) | \D0bar \rangle + \mbox{others} \biggr]
\Biggr \}
\nonumber
\end{eqnarray}
Here $\bar \psi \Gamma^\mu \psi = 
(\bar s \Gamma^\mu s - \bar d \Gamma^\mu d)$ and $\mbox{`others'}$
denotes operators with cyclic permutations of $\alpha, \mu, \mu$ indices.
$v$ represents the heavy quark velocity. Naive power counting argument 
of Ref.~\cite{Ge92} implies that the $U$-spin violating operator
$\bar \psi \Gamma^\mu \psi$ would scale like $m_s \Lambda^2$
and therefore, the overall contribution to $x$ and $y$ is multiplied by
a factor of $m_s^3$, compared to the leading term, where 
$x \sim m_s^4$ and $y \sim m_s^6$.
In order to develop an imaginary part (and so generate $y$), a gluon 
correction should be considered. Therefore, the contribution
of dimension nine operators to $y$ is suppressed by both $\alpha_s$ and 
phase space factors compared to $x$, 
$y^{(9)} \sim (\alpha_s / 16 \pi) x^{(9)} \ll x^{(9)}$. 
While it is impossible to estimate this contribution reliably
(there are unknown matrix elements of 15 operators), naive power
counting rules imply that it dominates the 
parametrically leading terms in the expansion of $x$~\cite{Ge92} and 
$y$~\cite{Bigi:2000wn}. 

The next important contribution to $y$ is obtained  
at the next order in $1/m_c$ and is given by a subset of matrix 
elements of the operators of dimension twelve.
This contribution is obtained by cutting all light fermion 
lines and adding a gluon to transfer large momentum. It is 
therefore represented by a set of eight-fermion operators.
While suppressed by $\alpha_s/m_c^2$, it again lifts another
factor of $m_s$. More importantly, $y^{(12)} \sim x^{(12)}$!  
This observation~\cite{Bigi:2000wn}
comes from the fact that imaginary part of the diagram that is
needed for generating $\DG$ can also be obtained
by dressing the gluon propagator by quark and gluon ``bubbles''.
The resulting $\alpha_s(m_c)$ suppression is largely compensated 
by the ``enhancement'' from the QCD $\beta$ function. 
This results in the estimate~\cite{Bigi:2000wn,GrFaLiPe}
\be
x,y \sim 0.1 \%,
\ee
which is obtained from the naive dimensional analysis, as
there are too many unknown matrix elements for the accurate
prediction to be made.

Indeed, the short-distance analysis, while systematic, is valid 
as long as one believes that the charmed quark is sufficiently 
heavy for $1/m_c$ expansion to be performed. 
Moreover, truly long-distance $SU(3)$ breaking effects might 
not be captured in the short distance analysis. For example, 
a contribution from a light quark resonance with 
$m_R \approx m_D$ would not be captured in this analysis. 
For a sufficiently narrow resonance, this provides a 
mechanism for breaking of local quark-hadron 
duality~\cite{Golowich:1998pz}. 

An alternative way of estimating $x$ and $y$ is to start 
from the long distance contributions generated by the
intermediate {\it hadronic} states. They arise from the
decays to intermediate states common to both $D^0$ and $\D0bar$.
Therefore, a sum over all possible $n$-particle intermediate
states allowed by the corresponding quantum numbers should be
taken into account in Eq.~(\ref{mixmatr}). In practice, only a 
few states are considered, so only an order-of-magnitude estimate
is possible. Even with this restriction, it is extremely difficult to
reliably determine the total effect from a given subset
of intermediate states due to the many decay modes with
unknown final state interaction (FSI) phases. In addition, 
hadronic intermediate states in $D^0 - \D0bar$ mixing are expected to
occur as $SU(3)$ flavor multiplets, so there are cancelations among
different contributions to $x$ and $y$ from the same multiplet.
These flavor $SU(3)$ relations can be analyzed. The initial 
$D$ state is an $SU(3)$ triplet, $D_i = (D^0,D^+,D^+_s)$, while the 
final state consists of a number of particles belonging 
to the octet representation,
\begin{equation}
M^i_k =
\left ( \begin{array}{ccc} 
\frac{\pi^0}{\sqrt{2}} + \frac{\eta}{\sqrt{6}}  & \pi^+ & K^+\\
\pi^- & -\frac{\pi^0}{\sqrt{2}} + \frac{\eta}{\sqrt{6}} & K^0 \\
K^- & \bar K^0 & -\sqrt{\frac{2}{3}} \eta
\end{array} \right )\ \ .
\end{equation}
A set of relations for the transition amplitudes 
$A_I = \langle D^0 | {\cal H}^{\Delta C=1}_W | I \rangle$
can be written. The effective Hamiltonian for $D$ transitions,
$ {\cal H}^{\Delta C=1}_W \sim (\bar \psi c)(\bar u \psi)$
with $\psi = s,d$ transforms as 
$\overline{15} \oplus 6 \oplus \bar 3 \oplus \bar 3$ 
under $SU(3)_F$.
Thus, $D_i$ and $M^i_k$ should be contracted 
with the vector $H(\bar 3)^i$ ($\bar 3$ Hamiltonian),
antisymmetric (wrt upper indices) tensor $H(6)^{ij}_k$ 
($6$ Hamiltonian), or symmetric tensor 
$H(\overline{15})^{ij}_k$ ($\overline{15}$ Hamiltonian).
The $SU(3)$ relations for $\DG$ follow as 
$\DG \sim \langle D^0 | {\cal H}^{\Delta C=1}_W | I \rangle 
\langle I | {\cal H}^{\Delta C=1}_W | \D0bar \rangle$ and are 
rather complicated for a generic multiparticle intermediate state.

Let us elaborate on the simplest possible contribution, 
due to intermediate single-particle states~\cite{Golowich:1998pz}. 
These are rather simple to analyze, as the number of such 
intermediate states is constrained.
A contribution to $y$ from a resonance state $R$
can be written as
\begin{equation}
y \bigg|_{res}=
\frac{1}{2 \Gamma m_D} \sum_{R} Im~
\frac{\langle D_2 | {\cal H}_W | R \rangle
\langle R | {\cal H}_W^\dagger | D_2 \rangle}
{m_D^2 - m_R^2 + i \Gamma_R m_D}
\ \ - \ \ (D_2 \leftrightarrow D_1) \ \ .
\label{deltaop}
\end{equation}
The pseudoscalar $0^{-+}$ (scalar $0^{++}$) intermediate states have
$CP = -1$ ($CP = +1$) and contribute (in the CP-limit) to the 
$D_1$ ($D_2$) part of the above equation. In principle, this 
contribution exhibits a resonant 
enhancement for a narrow resonance with $m_R \approx m_D$. 
In reality, light quark states with such large masses are not narrow.

In the limit of degenerate $s$ and $d$ quark masses the contribution 
from the entire $SU(3)$ multiplet would vanish, as expected from the 
GIM cancelation mechanism. Yet, $SU(3)$ is known to be badly broken in
$D$-decays~\cite{Falk:1999ts,cc94}, so a sizable value for the
width difference might not be surprising. 

A set of $SU(3)$ relations for the $D \to R$ transitions follow 
from the following transition amplitude
\begin{equation}
A(D \to R) = A_3 D_i M^i_k H(\bar 3)^k + A_6 D_i H(6)^{ik}_l M^l_k 
+ A_{15} D_i H(\overline{15})^{ik}_l M^l_k
\end{equation}
A contribution of the octet of pseudoscalar single-particle intermediate
states $\pi_{\rm H}$, $K_{\rm H}$, ${\bar K}_{\rm H}$, $\eta_{\rm H}$
(and possibly $\eta'_{\rm H}$ with $\eta_{\rm H}-\eta'_{\rm H}$
mixing angle $\theta_{\rm H}$) is
\begin{eqnarray}
y|^{\rm res}_{\rm octet} =
y^{(K_H)} - {1\over 4} y^{(\pi_H)}
- {3 \cos^2\theta_{\rm H} \over 4} y^{(\eta_H)}
- {1 \sin^2\theta_{\rm H} \over 4} y^{(\eta'_H)} \ \ ,
\label{spis}
\end{eqnarray}
with the mixing amplitudes induced by resonance $R$ calculated
to be
\begin{equation} \label{rescon1}
y^{\rm res} = - \frac{|H_R|^2}{m_D^3 \Gamma}
~{\gamma_R \over (1 - \mu_R)^2 + \gamma^2_R} \ \ ,
\label{delmg}
\end{equation}
where $|H_R|^2 \equiv \langle D^0 | {\cal H}_W | R \rangle
\langle R | {\cal H}_W^\dagger | \D0bar \rangle$, and
the dimensionless quantities
$\mu_R \equiv m_R^2/m_D^2$ and $\gamma_R \equiv
\Gamma_R/m_D$ are the reduced squared-mass and width
of the resonance.

No reliable information about the size of 
$\langle D | {\cal H}_W | R \rangle$ matrix elements
is available at the moment.
A typical contribution to $y$ from one 
$0^{-+}$ single-particle heavy intermediate state can be
calculated using vacuum insertion ansatz. 
This implies 
$|H_R|^2 = \mu_R f_R^2 m_D (G_F a_2 f_D \xi_d /\sqrt{2})^2$,
with $f_R$ being the resonance decay constant.
Making an ``educated guess'' about the size of $f_R$, it
can be shown that a typical contribution from 
a $0^{-+}$ amounts to $a~few \times 10^{-4}$
(see Ref.~\cite{Golowich:1998pz}),
but might be larger.

An estimate of $H_R$ for a $0^{++}$ single-particle
heavy intermediate state ( like $K^*(1430)$ or $K^*(1940)$)
can be obtained using the soft pion theorem arguments 
of Ref.~\cite{Gronau:1999zt} and measured branching
ratios for $D^+ \to R \pi^+$ transitions.
{\it Assuming} that 
expected corrections to the soft pion theorem are not 
large we derive for $R = K^*(1430)$
\begin{equation}
y_{0^{++}} = - \tan^2 \theta_C \frac{8 \pi f_\pi^2}{q_\pi}
\frac{{\cal B}(D^+ \to K^*(1430) \pi^+)}{f m_D^2}
\frac{\Gamma_{D^+}}{\Gamma_{D^0}}
\frac{\gamma_R}{(1-\mu_R)^2+\gamma_R^2},
\end{equation}
where $q_\pi=0.368~GeV$ is a pion's momentum,
${\cal B}(D^+ \to K^*(1430) \pi^+) \simeq 0.023$,
$f \equiv {\cal B}(K^*(1430) \to K \pi) \simeq 0.62$,
$f_\pi = 0.13~GeV$ is a pion's decay constant, and
$\Gamma_{D^+}/\Gamma_{D^0} \simeq 0.4$. This gives
\begin{equation}
|y_{0^{++}}(1430)| \simeq 0.02 \%,
\end{equation}
which is in the same ballpark as $y_{0^{-+}}$.
Now, if we {\it assume} that
$H_{K^*(1430)} \simeq H_{K^*(1940)}$, 
\begin{equation}
|y_{0^{++}}(1940)| \simeq 0.1 \%,
\end{equation}
It is clear from the Eq.({\ref{spis}) that $y=0$
in the $SU(3)_F$ limit, where $\mu_i = \mu_0, \gamma_i = \gamma_0$,
and $H_i = H_0$ for $i=\pi_H,K_H,\eta_H$ and
$\sin^2\theta_{\rm H} \to 0$.
It is therefore necessary to assess the pattern of
$SU(3)$-symmetry breaking in Eq.~(\ref{spis}). Neglecting
singlet-octet mixing and assuming that
\begin{eqnarray}
\mu_i &=& \mu_0 + \delta \mu_i, \nonumber \\
\gamma_i &=& \gamma_0 + \delta \gamma_i \\
|H_i| &=& |H_0| + \delta H_i, \nonumber
\end{eqnarray}
we obtain an estimate of $y$
\begin{eqnarray} \label{1psu3}
-y|^{\rm res}_{\rm octet} \times
m_D^3 \Gamma \frac{(1-\mu_0)^2+\gamma_0^2}{|H_0| \gamma_0}
&=&
2 \frac{\mu_0(1-\mu_0)}{(1-\mu_0)^2+\gamma_0^2}
\left[
\frac{\delta \mu_K}{\mu_0} -
\frac{1}{4} \frac{\delta \mu_\pi}{\mu_0} -
\frac{3}{4}\frac{\delta \mu_\eta}{\mu_0} \right]
\nonumber \\
&+& ~\frac{(1-\mu_0)^2 - \gamma_0^2}{(1-\mu_0)^2+\gamma_0^2}
\left[
\frac{\delta \gamma_K}{\gamma_0} -
\frac{1}{4} \frac{\delta \gamma_\pi}{\gamma_0} -
\frac{3}{4}\frac{\delta \gamma_\eta}{\gamma_0} \right]
\\
&+& 2 \left[
\frac{\delta H_K}{|H_0|} -
\frac{1}{4} \frac{\delta H_\pi}{|H_0|} -
\frac{3}{4}\frac{\delta H_\eta}{|H_0|} \right].
\nonumber
\end{eqnarray}
Unfortunately, many of the parameters of Eq.~(\ref{1psu3})
are not known. Yet, it's not unlikely that the total resonance
contribution could amount to $y \approx 0.1 \%$ or so.

Let us briefly discuss a contribution from charged pseudoscalar
{\it two-body} intermediate state. It was originally considered in
Refs.~\cite{dght,lw,Bu95} and estimated to be potentially large,
\begin{eqnarray} \label{2body}
y_2 =
\frac{1}{2m_D \Gamma} \sum_{p_1, p_2} Re \int
\frac{d^3 {\bf p}_1}{(2 \pi)^3 2 E_1}
 \frac{d^3 {\bf p}_2}{(2 \pi)^3 2 E_2}
\langle D^0 | {\cal H}_W | p_1, p_2 \rangle
\langle p_1, p_2 | {\cal H}_W^\dagger | \D0bar \rangle,
\end{eqnarray}
where one must sum over all intermediate state particles
$p_1, p_2$, not only ground state mesons. For the
charged pseudoscalar state $\{p_1, p_2\} = 
\{K^+,K^-\}$, $\{\pi^+,\pi^-\}$, $\{K^+,\pi^-\}$, 
and $\{K^-,\pi^+\}$. 
As before, the $SU(3)$ relations among amplitudes imply 
cancelations. These cancelations occur
within each multiplet, however broken $SU(3)$
assures that they are not complete. Residual contributions
from each multiplet then have to be summed up. 

In some cases available experimental data can be used.
For example, for $p_1, p_2 = K^+ K^-$ we easily
obtain from Eq.~(\ref{2body}) that
$y_{KK} = {\cal B}(D^0 \to K^+ K^-)$, which is well measured.
Thus, the charged pseudoscalar contribution can be easily
estimated
\begin{equation}
y_2 = (5.76 - 5.29 \cos \delta) \times 10^{-3},
\end{equation}
where the strong phase difference $\delta$ is defined in 
Eq.~(\ref{defphases}). Taking $-1 < \cos \delta < 0$ (see discussion 
in~\cite{Falk:1999ts,Bergmann:2000id}) implies that
\begin{equation}
0.6 \times 10^{-2} < y_2 < 1.1 \times 10^{-2}
\end{equation}
or
\begin{equation}
y_2 < 1.53 \times 10^{-3},
\end{equation}
if $\delta < 40^o$, as favored by hadronic models~\cite{Falk:1999ts}. 
Unfortunately, the experimental information about many 
other relevant hadronic decays is not available, so 
model-dependence of the final result is unavoidable.

We have to note, however, that phase space effects should 
profoundly distort the patterns of GIM cancelations for 
the intermediate states containing excited mesons~\cite{GrFaLiPe}. 
For example, let us take the decay modes with one ground state
and one excited state (first radial excitation) mesons, 
like $K(1460)$ or $\pi(1300)$.
Clearly, the final state $K(1460) K$ is kinematically 
forbidden, while other decays in the same $SU(3)$ multiplet
are not! Unfortunately, 
no experimental data exists for these transitions. 

To summarize our discussion, we note that it is quite likely
theoretically that $y \sim 0.1 \%$, as it is dominated by a 
SM $\Delta C =1$ contribution, whereas $x$ can be as
large as a percent in certain extensions of the Standard
Model. Some long-distance contributions to $y$ can also 
be as large as a percent, but they are either canceled 
by similar contribution form the same $SU(3)$ multiplet or
require values of strong phases that are unfavored by
$SU(3)$ and hadronic models.

\section{Experimental situation}

There are two intriguing experimental measurements providing
some information about $D \bar D$ mixing parameters.
The FOCUS experiment fits the time dependent decay rates of the 
singly-Cabibbo suppressed and the Cabibbo-favored modes to 
pure exponentials. We define $\hat\Gamma$ to be the parameter that is 
extracted in this way. More explicitly, for a time dependent decay rate with
$\Gamma[D(t)\rightarrow f]\propto e^{-\Gamma t}(1-z\Gamma t+\cdots)$, where
$|z|\ll1$, we have $\hat\Gamma(D\rightarrow f)=\Gamma(1+z)$.
The above equations imply the following relations:
\bea 
\hat\Gamma(D^0\rightarrow K^+K^-)\ &=&\
 \Gamma\ [1+R_m(y\cos\phi-x\sin\phi)],
\nonumber \\
\hat\Gamma(\barD\rightarrow K^+K^-)\ &=&\
 \Gamma\ [1+R_m^{-1}(y\cos\phi+x\sin\phi)],
\\
\hat\Gamma(D^0\rightarrow K^-\pi^+)\ &=&\
 \hat\Gamma(\barD\rightarrow K^+\pi^-)\ =\ \Gamma.
\nonumber
\eea
Note that deviations of $\hat\Gamma(D\rightarrow K^+K^-)$ from $\Gamma$ 
do not require that $y\neq0$. They can in principle be accounted 
for by $x\neq0$ and $\sin\phi\neq0$, but then they have a 
different sign in the $D^0$ and $\barD$ decays. FOCUS combines 
the two $D\rightarrow K^+K^-$ modes. To understand the 
consequences of such an analysis, one has to consider the relative
weight of $D^0$ and $\barD$ in the sample~\cite{Bergmann:2000id}. 
Let us define $A_{\rm prod}$ as the production asymmetry of 
$D^0$ and $\barD$, 
$A_{\rm prod}\equiv(N(D^0)-N(\barD))/(N(D^0)+N(\barD))$
Then, if $A_{\rm prod}$ is small (as suggested by E687 data) and
if $R_m^{\pm2}=1\pm A_m$, with $A_m$ being small (as suggested by CLEO), 
\bea \label{focustheory}
y_{\rm CP}\ &\equiv&\ {\hat\Gamma(D\rightarrow K^+K^-)\over
\hat\Gamma(D^0\rightarrow K^-\pi^+)}-1 
\nonumber \\
&=& y\cos\phi-x\sin\phi\left({A_m\over2}+A_{\rm prod}\right).
\eea
The one sigma range measured by FOCUS is
\be \label{focusex}
y_{\rm CP}=(3.42\pm1.57)\times10^{-2}.
\ee

The CLEO measurement gives the coefficient of each of the three 
terms ($1$, $\Gamma t$ and $(\Gamma t)^2$) in the doubly-Cabibbo suppressed
decays. Such measurements allow a fit to the parameters $R$, $R_m$, 
$x^\prime\sin\phi$, $y^\prime\cos\phi$, and $x^2+y^2$.  CLEO 
quotes the following one sigma ranges:
\bea \label{cleoex}
R\ &=&\ (0.48\pm0.13)\times10^{-2},
\nonumber \\
y^\prime\cos\phi\ &=&\ (-2.5^{+1.4}_{-1.6})\times10^{-2},
\\
x^\prime\ &=&\ (0.0\pm1.5)\times10^{-2},
\nonumber \\
A_m\ &=&\ 0.23^{+0.63}_{-0.80}.
\nonumber 
\eea
As we shall see shortly, a combination of FOCUS (\ref{focusex})
and CLEO (\ref{cleoex}) results provides powerful constraints
on the values of $D^0-\barD$ mixing parameters.

\section{Interpretation and Conclusions}

Let us now see the implications of the new CLEO and FOCUS 
measurements for the value of $y$. {\it We shall assume that the true
values of the mixing parameters are within one sigma of the results
provided by these two experiments}. First of all, based on the
available bounds on $x$, $\sin \phi$ and $|A_m|$~\cite{Bergmann:2000id},
one can argue that it is very unlikely that FOCUS result is 
accounted for by the second term in Eq.~(\ref{focustheory}). 
Therefore, {\it if the true values of the mixing parameters are 
within the one sigma ranges of CLEO and FOCUS measurements, 
then $y$ is of order of a (few) percent.} More specifically,
$y\cos\phi\approx0.034\pm0.016$! This is a rather surprising 
result (see Section 2). Also, if CLEO and FOCUS results are consistent,
then 
\bea \label{phase}
&& \cos\delta-(x/y)\sin\delta=-0.73\pm0.55,~ \mbox{or}
\nonumber \\
&& \cos\delta \sim  +0.65~ \mbox{if}~ |x|\sim|y|
\\
&& \cos\delta \sim -0.18~ \mbox{if}~ |x|\ll|y|
\nonumber
\eea
which leads to another interesting conclusion: 
{\it if the true values of the mixing parameters are within the 
one sigma ranges of CLEO and FOCUS measurements, then the 
difference in strong phases between the $D^0\rightarrow K^+\pi^-$ 
and $D^0\rightarrow K^-\pi^+$ decays is very large.} 

Since the strong phase $\delta$ vanishes in the $SU(3)$ flavor symmetry 
limit, the result of Eq.~(\ref{phase}) is also rather surprising
(for a discussion of the strong phase difference in $D \to K \pi$
see~\cite{Falk:1999ts,Petrov:1999iv}). On the other hand,
there are other known examples of $SU(3)$ breaking effects of 
order one in $D$ decays, so perhaps we should not be prejudiced 
against a very large $\delta$. 

Charm physics experiments have started to probe an interesting
region of $D^0 - \barD$ mixing parameter space, therefore new and
excited results from the existing and future~\cite{Petrov:1999fe} 
experiments are warranted.

This work was supported in part by the United States National Science
Foundation under Grant No.~PHY-9404057 and No.~PHY-9457916, 
and by the United States Department of Energy under Grant 
No.~DE-FG02-94ER40869.

\end{document}